\documentstyle[11pt,twoside,colloqOHP,epsfig]{article}
\begin{document}
\title{Multiplicity-Study of Exoplanet Host Stars} \author{M. Mugrauer$^1$,
R. Neuh\"auser$^{1}$, T.Mazeh$^{2}$, E. Guenther$^{3}$} \affil{$^1$Astrophysikalisches Institut,
Universit\"at Jena, Schillerg\"a{\ss}chen 2-3, 07745 Jena, Germany [markus@astro.uni-jena.de]}
\affil{$^3$Th\"uringer Landessternwarte Tautenburg, Sternwarte 5, 07778 Tautenburg, Germany
[guenther@tls-tautenburg.de] }\affil{$^2$ Tel Aviv University, Tel Aviv 69978, Israel
[mazeh@post.tau.ac.il]}
\begin{abstract}
We carry out a systematic search campaign for wide companions of exoplanet host stars to study
their multiplicity and its influence on the long-term stability and the orbital parameters of the
exoplanets. We have already found 6 wide companions, raising the number of confirmed binaries among
the exoplanet host stars to 20 systems. We have also searched for wide companions of Gl\,86, the
first known exoplanet host star with a white dwarf companion. Our Sofi/NTT observations are
sensitive to substellar companions with a minimum-mass of 35\,$\rm{M_{Jup}}$ and clearly rule out
further stellar companions with projected separations between 40 and 670\,AU.
\end{abstract}

\section{An imaging search campaign for wide companions of exoplanet host stars}

Some of the exoplanet host stars were found to be components of binary systems and first
statistical differences between exoplanets around single stars and exoplanets located in binary
systems were already reported by Zucker \& Mazeh (2002) as well as Eggenberger et al. (2004). In
particular, it seems that planets with orbital periods shorter than 40 days exhibit a difference in
their mass-period and eccentricity-period distribution.

However, all the derived statistical differences are based only on a small number of known binary
systems among the exoplanet host stars, i.e. their significance is sensitive to any changes in the
sample size. Furthermore in the statistical analyses it is assumed that most of the exoplanet host
stars are single star systems expect these stars known to be a component of a binary system.

In that context it is important to mention that the whole sample of exoplanet host stars was not
systematically surveyed so far for neither wide nor close companions, i.e. several more exoplanet
host stars, considered today as single stars, might be members of binary systems. Only search
campaigns for companions of the exoplanet host stars will clarify their multiplicity status and
will finally verify the significance of the reported statistical differences.

Therefore, we have started an imaging search program for wide visual companions of exoplanet host
stars, carried out with UFTI/UKIRT, SofI/NTT as well as MAGIC/CA~2.2m. We can find all directly
detectable stellar and substellar companions (m$\ga$40\,$\rm{M_{Jup}}$) with projected separations
from about 50 up to 1000\,AU. Thereby companions are identified first with astrometry (common
proper motion) and their companionship is confirmed with photometry and spectroscopy later on. So
far, 6 wide companions were detected, see Mugrauer et al. (2005a) for further details.

\section{Gl\,86\,B, a white dwarf companion of an exoplanet host star}

Queloz et al. (2000) reported a long-term linear trend in the radial velocity of the exoplanet host
star Gl\,86. Furthermore, after combining Hipparcos measurements with ground-based astrometric
catalogues, Jahrei\ss~(2001) showed that this star is a highly significant $\Delta\mu$ binary. Both
results point out that there should be a companion of stellar mass in orbit around Gl\,86. Els et
al. (2000) indeed detected a faint common proper motion companion, Gl\,86\,B, with a separation of
only $\sim$ 2\,arcsec and concluded that it is a late L or early T brown dwarf.

With NACO/SDI observations, Mugrauer \& Neuh\"auser (2005b) detected the orbital motion of this
companion which is the final proof that it is orbiting the exoplanet host star. Furthermore they
showed with IR spectroscopy data that Gl\,86\,B is a white dwarf, i.e. this companion is the causer
of the reported linear trends in the radial and astrometric motion of the exoplanet host star.
Gl\,86\,B is the first known white dwarf detected as a close companion of an exoplanet host star.
With their high contrast NACO/SDI imaging, Mugrauer \& Neuh\"auser (2005b) can already exclude
further stellar companions around Gl\,86 with projected separations between 1 and 23\,AU.

We present here further complementary observations of the Gl\,86 binary system, carried out in our
wide companion search program using SofI/NTT. With these observations, we can clearly rule out
additional wide stellar companions around Gl\,86 with projected separations between 40 and 670\,AU
(see Fig.\,1).

\begin{figure}[tb]
\plotone{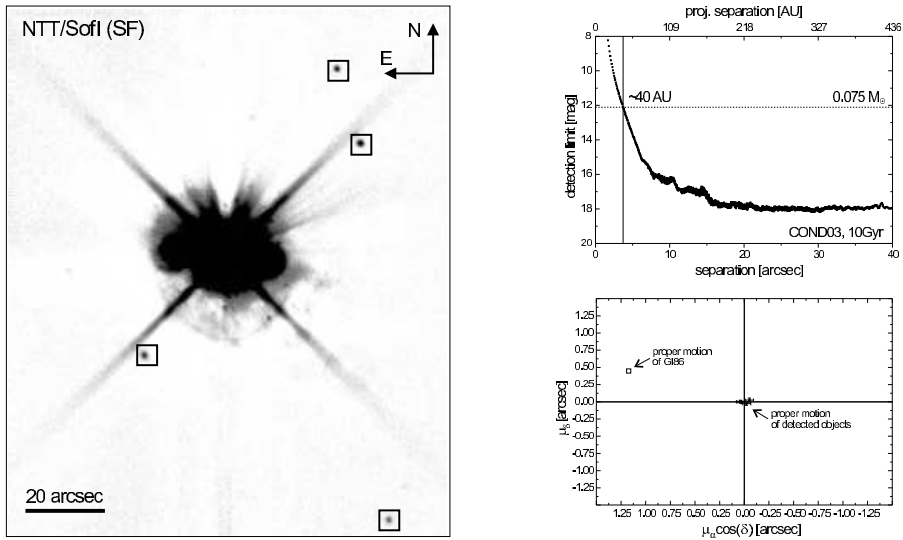}\caption{The left panel shows the 1st epoch H-band image of Gl\,86 obtained
with SofI/NTT in Dec. 2002. We observed the star again in 2nd epoch in June 2003. A detection limit
of H=18\,mag (S/N=10) is reached and substellar companions with a minimum-mass of
35\,$\rm{M_{Jup}}$ are detectable (see right upper plot). The proper motion between the 1st and 2nd
epoch imaging of all detected objects is illustrated in the right lower diagram.}
\end{figure}

\end{document}